\documentclass[journal]{IEEEtran}
\newcommand{\bunderline}[1]{\underline{#1\mkern-4mu}\mkern4mu }

\usepackage{cite}
\newcommand{\ubar}[1]{\mkern2mu\underline{\mkern-2mu #1\mkern-2mu}\mkern2mu}

\ifCLASSINFOpdf
   \usepackage[pdftex]{graphicx}
   \usepackage{epstopdf}
   \DeclareGraphicsExtensions{.pdf,.jpeg,.png}
\else
\fi
\usepackage[cmex10]{amsmath}
\usepackage{amssymb}
\usepackage{array}
\begin{document}
\title{A Simple Lower Bound on the Noncoherent Capacity of Highly Underspread Fading Channels}

\author{Steven~Herbert,~\IEEEmembership{Student~Member,~IEEE,}
        Ian~Wassell,
        Tian-Hong~Loh,~\IEEEmembership{Member,~IEEE,}% <-this % stops a space
\thanks{This work is supported by the U.K. Engineering and Physical Sciences Research Council (EPSRC) and National Physical Laboratory (NPL) under an EPSRC-NPL Industrial CASE studentship programme on the subject of Intra-Vehicular Wireless Sensor Networks. The work of T. H. Loh was supported by the 2009–-2012 Physical Program and 2012–-2015 Electromagnetic Metrology Program of the National Measurement Office, an Executive Agency of the U.K. Department for Business, Innovation and Skills, under Projects 113860 and EMT13020, respectively.}        
\thanks{Steven Herbert is with the Computer Laboratory, University of Cambridge, CB3 0FD, UK, and the National Physical Laboratory, TW11 0LW, UK. Contact: sjh227@cam.ac.uk}% <-this % stops a space
\thanks{Ian Wassell is with the Computer Laboratory, University of Cambridge, CB3 0FD, UK. Contact: ijw24@cam.ac.uk}
\thanks{Tian Hong Loh is with the National Physical Laboratory, TW11 0LW, UK. Contact: tian.loh@npl.co.uk}
}
\maketitle

\begin{abstract}
%\boldmath
Communication channels are said to be underspread if their coherence time is greater than their delay spread. In such cases it can be shown that in the infinite bandwidth limit the information capacity tends to that of a channel with perfect receiver Channel State Information (CSI). This paper presents a lower bound on the capacity of a channel with finite bandwidth, expressed in a form which is mathematically elegant, and computationally simple. The bounding method exploits the fact that most actual channels are highly underspread; and that typically more is known about their impulse response than the channel time variation. The capacity is lower bounded by finding an achievable rate for individual time blocks which are shorter than the channel coherence time, in an orthogonal frequency division multiplexing system model.\\
A highly underspread channel of particular interest is the in-vehicle channel, and a numerical example is given to verify that the capacity is indeed approximately that of a channel with perfect receiver CSI. The resulting lower bound is shown to be tighter than those previously derived.
\end{abstract}


% \fi
\IEEEpeerreviewmaketitle

\section{Introduction}
\label{int}
\IEEEPARstart{A}{ctual} communication channels are typically highly underspread: their delay spread is much smaller than their coherence time. A more exact definition can be made by describing the action of a wireless channel as a linear operator $\mathbb{H}: \mathcal{L}^2 \to \mathcal{L}^2$. The action of $\mathbb{H}$ can be expressed in terms of the scattering function, $\mathcal{C}_\mathbb{H}(\nu,\tau)$, where $\nu$ is the Doppler shift, and $\tau$ is the time delay \cite{bello}. For a Wide-Sense Stationary Uncorrelated Scattering (WSSUS) channel the non-zero region of the scattering function is defined: $\mathcal{C}_\mathbb{H}(\nu,\tau) =0$ for all $(\nu,\tau) \notin [-\nu_0, \nu_0] \times [-\tau_0,\tau_0]$. Letting $\Delta_\mathbb{H} = 4 \nu_0 \tau_0 $, the channel is said to be underspread if $\Delta_\mathbb{H} < 1$ \cite{durisi1}. For land-mobile channels $\Delta_\mathbb{H} \approx 10^{-3}$, for indoor channels $\Delta_\mathbb{H} \approx 10^{-7}$ \cite{durisi1}, and for in-vehicle channels $\Delta_\mathbb{H}$ is typically of the order $10^{-5}$ \cite{us2}. Durisi \textit{et al} argue that the WSSUS model is appropriate for many scenarios and our previous work shows that, for in-vehicle channels (which are of particular interest \cite{wsnv} to us), the channel can be assumed to be wide-sense stationary [\cite{us2}~Assumption~1] and to have uncorrelated scattering [\cite{us1}~Assumptions~1,2].\\

As identified by Durisi \textit{et al} \cite{durisi3}, the noncoherent capacity of underspread fading channels has been the subject of research for a long time, with early work typically focussing on characterising the noncoherent capacity in the infinite bandwidth limit \cite{gallager, ken69, pie66, vit67}. It is shown that in this situation, the capacity tends to that of an Additive White Gaussian Noise (AWGN) channel with perfect Channel State Information (CSI) available at the receiver, as derived by Shannon \cite{shannon1948}. Durisi \textit{et al} themselves use this as motivation to further investigate the noncoherent capacity of underspread channels from a more general starting point. This they achieve by transmitting with symbols which are well localised in both time and frequency and subsequently deriving a number of lower and upper bounds on the channel capacity, which have been optimised either for the low bandwidth or high bandwidth regimes \cite{durisi1, durisi2, durisi3}. It is also important to note that, in these papers, the capacity allows for a constraint on the peak power in frequency and time, which has been included to take into account practical considerations of actual radio transmitting and receiving equipment, and maximum power regulations.\\

In this paper, we contend that an alternative method of lower bounding can add some more important insights to the field of the noncoherent capacity of underspread channels. Our method exploits the intuitive property that, for highly underspread channels, the channel remains unchanged for a time duration (the channel coherence time) which is much greater than its delay spread, and therefore it should be possible to learn CSI within one coherence time interval. This is achieved by using an Orthogonal Frequency Division Multiplexing (OFDM) scheme \cite{tse1, goldsmith} to define the input as a vector which modulates the channel as discrete frequencies spread with constant intervals. Noticing that for highly underspread channels, the frequency response will be highly correlated at these discrete frequencies, the correlation between successive input and output pairs can be used to infer CSI.

\subsection{Contributions}

This approach yields a lower bound which provides a number of specific contributions to complement that which has previously been achieved in the field of the noncoherent capacity of underspread channels:

\begin{enumerate}
\item The starting assumptions for the bound are that both the transmitter and the receiver have knowledge of the coherence time of the channel and of the statistical impulse response, but not the specific realisation. This constitutes a more general assumption compared to that of Durisi \textit{et al} \cite{durisi1, durisi2, durisi3} where it was assumed that the transmitter and receiver have knowledge of the statistical scattering function (but not its realisation).
\item The bounding method can be understood intuitively, and is computationally simple. Moreover, the lower bound deals with the bandwidth in a very clear manner, and it is straightforward to evaluate the bandwidth sufficient to achieve a specified fraction of the AWGN capacity.
\item There is evidence that, at least for some channels, the lower bound presented here is tighter than those proposed by Durisi \textit{et al} \cite{durisi1, durisi2, durisi3}.\\
\end{enumerate}

Furthermore, our bounding method reduces to a lower bound on a channel where the channel responses form a multivariate Gaussian. It is shown that the Markov case, where each channel response relies only on the previous response, lower bounds the general case, and we contend that this result may have much wider application than merely the noncoherent capacity of underspread fading channels.

\subsection{Paper Organisation}

In Section~\ref{chmod1} a general channel model is defined in terms of its frequency response and in Section~\ref{chmod2} it is explained how this can be used in an OFDM scheme to evaluate a lower bound on the channel capacity. In Section~\ref{lb} the main results are presented (with some parts of the proof in the appendices) in the form of the aforementioned lower bound, whilst in Section~\ref{ex} a numerical worked example is given and finally in Section~\ref{conc} conclusions are drawn.

\subsection{Notation}
\label{notation}

Italicised and non-italicised symbols are used for frequency and time domain variables respectively. Scalars are non-bold lower case, as in general are functions (i.e., $\mathrm{x}$ for the time domain, $x$ for the frequency domain), vectors are bold lower-case (i.e., $\mathbf{x}$ for the time domain, $\boldsymbol{x}$ for the frequency domain), a single element from a vector or matrix is non-bold lower-case with subscript to denote its index (i.e., $\mathrm{x_i}$ for a vector and $\mathrm{x_{i,j}}$ for a matrix in the time domain; and $x_i$ for a vector and $x_{i,j}$ for a matrix in the frequency domain), truncated vectors are bold lower-case with subscript to denote first element and superscript to denote final element (i.e., $\mathbf{x_i^j}$ for the time domain, $\boldsymbol{x_i^j}$ for the frequency domain) and matrices are upper-case (i.e., $\mathrm{X}$ for the time domain, $X$ for the frequency domain).\\

Convolution is denoted $\ast$, $(.)^*$ is used to denote complex conjugation and $(.)^T$ to denote the transpose. $\mathcal{N}$ denotes the normal distribution, $\mathcal{CN}$ denotes the complex normal distribution, $\mathcal{FT}$ denotes the Fourier transform and $\odot$ denotes the \textit{Hadamard} (element-wise) product. The magnitude of a complex number is denoted $|.|$, as is the determinant of a matrix, however it is always clear in context which is meant.\\

Finally, it is convenient to represent complex numbers as vectors, and when multiplied together as a matrix acting on a vector. Letting $x$ be a number, which in general may be complex:
\begin{IEEEeqnarray}{rcl}
\bunderline{x} & = & \begin{bmatrix} \textnormal{Re}(x) \\ \textnormal{Im}(x) \end{bmatrix} , \nonumber \\
\bunderline{X} & = & \begin{bmatrix} \textnormal{Re}(x) & \textnormal{-Im}(x) \\ \textnormal{Im}(x) & \textnormal{Re}(x) \end{bmatrix} , \nonumber 
\end{IEEEeqnarray}
for example (letting $z$ also be a number which may in general be complex):
\begin{equation}
z \times x = \bunderline{Z} \bunderline{x} = \bunderline{X} \bunderline{z}. \nonumber
\end{equation}

The notation can also be generalised to complex vectors. Letting $\boldsymbol{x}$ be a vector of size $n$ with each element in general a complex number:
\begin{IEEEeqnarray}{rcl}
\boldsymbol{\bunderline{x}} & = & \begin{bmatrix} \bunderline{x}_1 \\ \bunderline{x}_2 \\ \vdots \\ \bunderline{x}_n \end{bmatrix} , \nonumber \\
\boldsymbol{\bunderline{X}} & = & \begin{bmatrix} \bunderline{X}_1 & \ubar{0} & \ubar{0} & \cdots \\   \ubar{0} & \bunderline{X}_2 & \ubar{0} & \cdots \\  \ubar{0} & \ubar{0} & \ddots & \\ \vdots & \vdots & & \bunderline{X}_n \end{bmatrix}. \nonumber
\end{IEEEeqnarray}

\section{Channel frequency response}
\label{chmod1}

Let $\mathrm{P}_{\mathbb{H}}(\tau)$ be the instantaneous channel Power Delay Profile (PDP), from which a truncated version is defined:
\begin{equation}
\label{fr6}
\mathrm{P}_{\mathbb{H}}'(\tau) =  \Bigg\{ \begin{array}{ll}  \mathrm{P}_{\mathbb{H}}(\tau) & \textnormal{if} \,\, 0 \leq \tau < \tau_t  \\ 0 & \textnormal{otherwise} . \end{array}
\end{equation}
where $\tau_t$ is chosen such that the error between the truncated PDP and the original PDP is small, and this error will later be treated as additive noise.\\

The bounding method requires information regarding the in-vehicle channel characterisation in the frequency domain. At a randomly chosen frequency, with time period which is short compared to the delay spread of the signal, the distribution of the phase of the various multipath components will be uniform, and thus the frequency response will be a Zero Mean Circularly Symmetric (ZMCS) Gaussian random variable; defining this as $z(\omega)$, let:
\begin{equation}
\label{fr10}
\bunderline{z}(\omega) \sim \mathcal{N}\left(\bunderline{z}(\omega); \ubar{0}, \Sigma_z \right),
\end{equation}
where $\mathcal{N}$ is the Gaussian distribution and $\omega$ is angular frequency (i.e., $\omega = 2\pi f$ where $f$ is frequency) :
\begin{equation}
\label{fr20}
\Sigma_z = \begin{bmatrix} \sigma^2_z & 0 \\ 0 & \sigma^2_z \end{bmatrix},
\end{equation}
where $\sigma^2_z$ is the variance.\\

The bounding method also requires the conditional distribution of the frequency response, given the frequency response at a known separation, $(\Delta \omega)$, i.e., $P(z(\omega)|z(\omega - \Delta \omega))$.

\subsection*{Proposition 1:}

For the channel with PDP defined in (\ref{fr6}), the conditional distribution of the frequency response, given the frequency response at a known separation can be expressed:
\begin{equation}
\label{fr30}
P(\bunderline{z}(\omega) | \bunderline{z}(\omega- \Delta \omega))  = \mathcal{N}\left( \bunderline{z}(\omega); \bunderline{\mu}_{a} , \Sigma_{a} \right),
\end{equation}
where:
\begin{IEEEeqnarray}{rcl}
\label{fr40}
\bunderline{\mu}_{a}  & \, = \, & \bunderline{A} \bunderline{z}_{(\omega- \Delta \omega)}, \\
\label{fr50}
\Sigma_{a} & \, = \, &  \sigma_z^2 \begin{bmatrix} 1-|a|^2 & 0 \\ 0 & 1-|a|^2 \end{bmatrix},
\end{IEEEeqnarray}
where $z_{(\omega- \Delta \omega)}$ is the realisation of $z(\omega- \Delta \omega)$, $\bunderline{A}$ is the matrix version of the complex number $a$ (i.e., according to the notation in (\ref{notation})) and:
\begin{equation}
\label{fr60}
a = \frac{\int_0^{\tau_t} \mathrm{P}'_\mathbb{H} (\tau) e^{-j \Delta \omega \tau} \, \mathrm{d} \tau }{\int_0^{\tau_t} \mathrm{P}'_\mathbb{H} (\tau) \, \mathrm{d} \tau } .
\end{equation}

\subsection*{Proof:}

Consider splitting $\mathrm{P}_{\mathbb{H}}'(\tau)$ into an integer number of time intervals each of duration $\Delta \tau$. Assuming that in each time interval there are many arriving rays, and that the frequency is sufficiently high such that the phase of each arriving ray can be considered to be a random variable drawn from a uniform distribution, then the resultant signal from each time interval is a ZMCS complex Gaussian random variable. Each of these will be independent, given the uncorrelated scattering assumption, which is a necessary part of the WSSUS assumption stated in Section~\ref{int}. The joint distribution of the signal from these intervals can thus be expressed as a multivariate complex Gaussian distribution: The discrete signal vector, $\mathbf{z}$, is of size $\mathrm{K}$ where $\mathrm{K} = \tau_t / \Delta \tau$ and the $\mathrm{k}^{th}$ element occurs at $\tau=\mathrm{k} \Delta \tau$:
\begin{IEEEeqnarray}{rcl}
\label{ch10b}
\mathbf{z} & \, \sim \, & \mathcal{C}\mathcal{N} (\mathbf{z}; \boldsymbol{0}, \Gamma, 0) , \\
\Gamma & \, = \, &  \mathbb{E} \left( \mathbf{z} (\mathbf{z}^{*T} ) \right) , \nonumber \\
\label{ch20b}
& \, = \, & \textnormal{diag}(2\sigma^2_\mathrm{k})
\end{IEEEeqnarray}
where $\mathcal{C}\mathcal{N}$ is the complex Gaussian distribution, and:
\begin{equation}
\label{ch30b}
\sigma^2_\mathrm{k} = \left(\mathrm{P}_{\mathbb{H}}'(\mathrm{k} \Delta\tau)\right) \Delta \tau.
\end{equation}

By a Fourier transform, (\ref{ch10b}) can be used to approximate the frequency response at $\omega$ and $(\omega - \Delta \omega)$, for sufficiently small values of $\Delta \tau$:
\begin{IEEEeqnarray}{rcl}
P\left(\begin{bmatrix} z(\omega) \\ z(\omega- \Delta \omega) \end{bmatrix} \right) & \approx & \mathcal{C}\mathcal{N} \Bigg(\begin{bmatrix} z(\omega) \\ z(\omega- \Delta \omega) \end{bmatrix}; \nonumber \\
\label{ch40b}
& & \, \, \, \, \, \, \, \, \, \, \, \, \, \, \, \, \,
\boldsymbol{0},  \begin{bmatrix} \boldsymbol{f} \\ \boldsymbol{g} \end{bmatrix} \Gamma \begin{bmatrix} \boldsymbol{f} \\ \boldsymbol{g} \end{bmatrix}^{*T}, 0\Bigg)
\end{IEEEeqnarray}
where $\boldsymbol{f}$ and $\boldsymbol{g}$ are row vectors, each of size $\mathrm{K}$, with $\mathrm{k}^{th}$ elements:
\begin{IEEEeqnarray}{rcl}
\label{ch60b}
f_\mathrm{k} & \, = \, & e^{-j \omega \mathrm{k} \Delta \tau}, \\
\label{ch70b}
g_\mathrm{k}  & \, = \, &  e^{-j (\omega- \Delta \omega) \mathrm{k} \Delta \tau} .
\end{IEEEeqnarray}

Let:
\begin{equation}
\label{ch80b}
\begin{bmatrix} \boldsymbol{f} \\ \boldsymbol{g} \end{bmatrix} \Gamma \begin{bmatrix} \boldsymbol{f} \\ \boldsymbol{g} \end{bmatrix}^{*T} = \begin{bmatrix} \gamma_1 & \gamma_2 \\ \gamma_3 & \gamma_4 \end{bmatrix},
\end{equation}
then, noting that $\tau=\mathrm{k} \Delta \tau$:
\begin{IEEEeqnarray}{rcl}
\label{ch90b}
\gamma_1 & \, = \, & \sum_{\mathrm{k}=0}^{\mathrm{K}-1} e^{-j \omega \tau}  \left( 2 \mathrm{P}_{\mathbb{H}}'(\tau) \Delta \tau \right) e^{j \omega \tau} , \\
\label{ch100b}
\gamma_2 & \, = \, & \sum_{\mathrm{k}=0}^{\mathrm{K}-1} e^{-j \omega \tau} \left( 2 \mathrm{P}_{\mathbb{H}}'(\tau) \Delta \tau \right) e^{j (\omega- \Delta \omega) \tau}, \\
\label{ch110b}
\gamma_3 & \, = \, & \sum_{\mathrm{k}=0}^{\mathrm{K}-1} e^{-j (\omega-\Delta \omega) \tau} \left( 2 \mathrm{P}_{\mathbb{H}}'(\tau) \Delta \tau \right) e^{j \omega \tau}, \\
\label{ch120b}
\gamma_4 & \, = \, & \sum_{\mathrm{k}=0}^{\mathrm{K}-1} e^{-j (\omega-\Delta \omega) \tau} \left( 2 \mathrm{P}_{\mathbb{H}}'(\tau) \Delta \tau \right) e^{-j (\omega-\Delta \omega) \tau}. 
\end{IEEEeqnarray}
Let: $\Delta \tau \to 0$ (and thus adjusting $\mathrm{K}$ such that $\tau_t$ does not vary):
\begin{IEEEeqnarray}{rcccl}
\label{ch130b}
\gamma_1 & \, = \, & \gamma_4 & \, = \, &  \int_0^\infty 2 \mathrm{P}_{\mathbb{H}}'(\tau) \, \mathrm{d} \tau, \\
\label{ch140b}
&& \gamma_2 & \, = \, & \int_0^\infty 2 \mathrm{P}_{\mathbb{H}}'(\tau)  e^{-j \tau \Delta \omega}  \, \mathrm{d}\tau, \\
\label{ch150b}
&& \gamma_3 & \, = \, & \int_0^\infty 2 \mathrm{P}_{\mathbb{H}}'(\tau)  e^{j \tau \Delta \omega}  \, \mathrm{d}\tau.
\end{IEEEeqnarray}

Noticing that $[z(\omega); z(\omega- \Delta \omega)]$ is ZMCS complex Gaussian, it can be expressed as a multivariate Gaussian:
\begin{equation}
\label{ch170b}
\begin{bmatrix} \bunderline{z}(\omega) \\ \bunderline{z}(\omega - \Delta \omega) \end{bmatrix} \sim \mathcal{N}\left(\begin{bmatrix} \bunderline{z}(\omega) \\ \bunderline{z}(\omega - \Delta \omega) \end{bmatrix} ; \boldsymbol{\ubar{0}}, \Sigma_{\textnormal{t}}\right),
\end{equation}
where:
\begin{equation}
\label{ch180b}
\Sigma_{t} = \sigma_z^2 \begin{bmatrix}  1 & 0 & \textnormal{Re}(a) & -\textnormal{Im}(a) \\ 0 & 1 & \textnormal{Im}(a) & \textnormal{Re}(a) \\ \textnormal{Re}(a) & \textnormal{Im}(a) & 1 & 0 \\ -\textnormal{Im}(a) & \textnormal{Re}(a) & 0 & 1 \end{bmatrix} ,
\end{equation}
and, by definition:
\begin{IEEEeqnarray}{rcl}
\label{fr100}
\sigma^2_z & \, = \, & \int_0^\infty  \mathrm{P}_{\mathbb{H}}'(\tau) \, \mathrm{d} \tau, \\
\label{fr110}
a & \,  = \, & \frac{\int_0^{\tau_t} \mathrm{P}'_\mathbb{H} (\tau) e^{-j \Delta \omega \tau} \, \mathrm{d} \tau }{\int_0^{\tau_t} \mathrm{P}'_\mathbb{H} (\tau) \, \mathrm{d} \tau }.
\end{IEEEeqnarray}

The conditional distribution of $(z(\omega) | z(\omega- \Delta \omega))$ can also be found, letting the realisation of $z(\omega- \Delta \omega) = z_{\omega-\Delta \omega}$:
\begin{equation}
\label{ch240b}
(z(\omega) | z(\omega- \Delta \omega))  \sim \mathcal{N}(\bunderline{\mu}_{a} , \Sigma_{a}),
\end{equation}
where:
\begin{equation}
\label{ch250b}
\bunderline{\mu}_{a}  = \bunderline{A} \bunderline{z}_{\omega-\Delta \omega}
\end{equation}
and:
\begin{equation}
\label{ch260b}
\Sigma_{a}  =  \sigma_z^2 \begin{bmatrix} 1-|a|^2 & 0 \\ 0 & 1-|a|^2 \end{bmatrix},
\end{equation}
thus proving Proposition~1.

\section{Using orthogonal frequency division multiplexing to find an achievable rate}
\label{chmod2}

Consider a block fading model \cite{tse1, goldsmith}, where the continuously varying channel is split into blocks in which the fading is treated as identical, i.e., the channel remains virtually unchanged within one block. To find an achievable rate, an OFDM \cite{tse1, goldsmith} scheme is used. Owing to the fact that the channel is highly underspread, the block length can be chosen such that it is much shorter than the truncated version of the PDP, $\textnormal{P}'_\mathbb{H}(\tau)$.\\

The assumption that the channel has the Wide-Sense Stationary Uncorrelated Scattering property means that, without loss of generality, the energy associated with the part of the channel impulse response which does vary during a block, and that owing to the part of the impulse response removed during the truncation of $\textnormal{P}_\mathbb{H}(\tau)$ to form $\textnormal{P}'_\mathbb{H}(\tau)$, can both be treated as AWGN.\\

The channel is therefore split into blocks of length, $T_B$, each of which has a cyclic prefix of length, $T_t \geq \tau_t$, which can be chosen to be negligible compared to the block length (again due to the highly underspread property). $T_B$ and $T_t$ must be chosen such that $WT_B$ and $WT_t$ are integers (where $W$ is the bandwidth, and $N=WT_B$ is the total number of subcarriers).\\  

According to the Sampling Theorem \cite{shannon1948, gallager}, the waveform in one block can be reconstructed from samples spaced $1/2W$~s apart. Performing an Inverse Discrete Fourier Transform (IDFT) on the resulting vector of samples (i.e., in the time domain) yields a vector of samples in the frequency domain. These are spaced $1/T_B$~Hz apart. Choosing to define the input signal in the frequency domain, the channel output can be expressed:
\begin{equation}
\label{ofdmeq10}
\boldsymbol{y} = \boldsymbol{z} \odot \boldsymbol{x} + \boldsymbol{n},
\end{equation}
where $\boldsymbol{y}$ is a vector of outputs, $\boldsymbol{z}$ is a vector of the channel frequency response, $\boldsymbol{x}$ is a vector of the channel symbols, $\boldsymbol{n}$ is a vector of AWGN samples and $\odot$ denotes element-wise multiplication. All these vectors are of size $N = WT_B$. Noting that all the elements of the vectors are complex then (\ref{ofdmeq10}) can be expressed:
\begin{equation}
\label{ofdmeq20}
\boldsymbol{\bunderline{y}} = \bunderline{Z} \boldsymbol{\bunderline{x}} + \boldsymbol{\bunderline{n}}.
\end{equation}

Noting that the cyclic prefix allows cyclic convolution to be treated as linear convolution (i.e., as if the channel were LTI), the channel frequency response can be expressed:
\begin{IEEEeqnarray}{rcl}
\label{ofdmeq30}
P(\bunderline{z}_i) & \, = \, & \mathcal{N}\left(\bunderline{z}_i; \ubar{0}, \Sigma_z \right), \\
\label{ofdmeq40}
P(\bunderline{z}_i | \bunderline{z}_{i-1}) & \,  = \, & \mathcal{N}\left( \bunderline{z}_i; \bunderline{\mu}_{a} , \Sigma_{a} \right),
\end{IEEEeqnarray}
where:
\begin{IEEEeqnarray}{rcl}
\label{ofdmeq50}
\bunderline{\mu}_{a}  & \, = \, & \bunderline{A} \bunderline{z}_{i-1}, \\
\label{ofdmeq60}
\Sigma_{a} & \, = \, &  \sigma_z^2 \begin{bmatrix} 1-|a|^2 & 0 \\ 0 & 1-|a|^2 \end{bmatrix},
\end{IEEEeqnarray}
with $a$ as defined in (\ref{fr60}). When the channel is highly underspread it will be the case that $a \approx 1$ and therefore $\Sigma_{a} \approx 0$. Thus there is little variation in the channel frequency response given the previous frequency response. This is to be expected, and it is this property which allows the lower bound to be found.

\section{A lower bound on the channel capacity}
\label{lb}

There is a non-zero probability that for any given time block, the channel will be in outage, however considering the channel as a whole (i.e., across several independent time blocks), an achievable rate, $\mathrm{R}$, can be defined using the Channel Coding Theorem \cite{gallager}:
\begin{equation}
\label{m10}
\mathrm{R}  \geq  \frac{1}{N} \mathcal{I}(\boldsymbol{x}_0^{N-1};\boldsymbol{y}_0^{N-1}) \, \, \textnormal{bit}\, \textnormal{s}^{-1}\, \textnormal{Hz}^{-1}.
\end{equation}
subject to:
\begin{equation}
\label{m20}
\mathrm{P}_{\textnormal{ave}} \geq \lim_{\substack{N' \to \infty}} \frac{1}{N'} \sum_{i=0}^{N'-1} |x_i|^2,
\end{equation}
where $\mathcal{I}(.;.)$ is mutual information and $\mathrm{P}_{\textnormal{ave}}$ is the average power constraint. Note that this average is defined across an infinite number of input symbols, $x_i$, and thus an infinite number of time blocks.

The input distribution on $x$ is chosen to be a series of Independent Identically Distributed (IID) ZMCS Gaussian random variables:
\begin{equation}
\label{m30}
\bunderline{x}_i \sim  \mathcal{N} (\bunderline{x}_i ; \ubar{0}, \Sigma_x),
%\begin{bmatrix} \textnormal{Re}(x_i) \\ \textnormal{Im}(x_i) \end{bmatrix}   \sim  \mathcal{N} (\mathbf{0}, \Sigma_x),
\end{equation}
where:
\begin{equation}
\label{m40}
\Sigma_x  =  \begin{bmatrix} \sigma_x^2 & 0  \\ 0 & \sigma_x^2 \end{bmatrix} .
\end{equation}

Likewise, the additive white noise is modelled as IID ZMCS Gaussian random variables:
\begin{equation}
\label{m50}
\bunderline{n}_i \sim  \mathcal{N} (\bunderline{n}_i ; \ubar{0}, \Sigma_n),
\end{equation}
where:
\begin{equation}
\label{m60}
\Sigma_n  =  \begin{bmatrix} \sigma_n^2 & 0  \\ 0 & \sigma_n^2 \end{bmatrix} .
\end{equation}

\subsection{Bounding idea}

The aim is to lower bound (\ref{m10}). In essence, the bounding method is similar to a \textit{Kalman} filter \cite{kalman}, where the `state' is the channel response at the discrete frequencies corresponding to the input (i.e., where successive frequency responses are correlated) and the noisy measurement is formed by the information signal input-output pair. For each successive discrete frequency response, some CSI is learned, and thus eventually (assuming the frequency separation between discrete frequency responses is small) the CSI approaches perfect CSI.\\

There is no feedback in the system, i.e., because successive values of $x_i$ have no dependence on previous values of $y_i$, $z_i$ or $n_i$. This property is used throughout the bounding process to simplify various expressions.

\subsection*{Theorem 2:}

There exists a lower bound, $\mathrm{L}_1$, on the achievable rate:

\begin{equation}
\label{m70}
\mathrm{R} \geq \mathrm{L}_1 =  \frac{1}{N} \sum_{i=0}^{N-1} \mathcal{I}_i \, \, \textnormal{bit}\, \textnormal{s}^{-1}\, \textnormal{Hz}^{-1},
\end{equation}
where:
\begin{equation}
\label{m80}
\mathcal{I}_i = \mathbb{E} \left( \log_2 \left(\frac{|\mu'_i|^2 \sigma_x^2 + \sigma_n^2}{|x_i|^2 \sigma_i^2 +\sigma_n^2} \right) \right),
\end{equation}
and:
\begin{IEEEeqnarray}{rcll}
%\sigma^2_i & \, = \, & \sigma^2_z, & \,\,\,\,\, i=0 \\
\label{m90}
\sigma^2_i & \, = \, & \Bigg\{ \begin{array}{ll} \sigma^2_z & \textnormal{if} \,\,i=0, \\ \begin{array}{ll} (1-|a|^2)\sigma^2_z  \\ + |a|^2(\sigma_{i-1}^{-2}+|x_{i-1}|^{2} \sigma_{n}^{-2})^{-1} \end{array} & \textnormal{if} \,\,i>0, \end{array} \\
\label{m100}
\bunderline{\mu}'_i & \, \sim \, & \mathcal{N}\left( \bunderline{\mu}'_i ;\ubar{0}, \begin{bmatrix} \sigma_z^2 - \sigma_i^2 & 0 \\ 0 & \sigma_z^2 - \sigma_i^2 \end{bmatrix} \right).
\end{IEEEeqnarray}

The terms $\sigma^2_i$ and $\bunderline{\mu}'_i$ represent the variance and mean of the estimate of the $i^{\textnormal{th}}$ frequency response respectively. The term $\sigma^2_i$ generally decreases with $i$, and notice that as $\sigma^2_i \to 0$:
\begin{equation}
\label{m280}
\mathcal{I}_i \to \mathbb{E} \left( \log_2 \left( 1 + |z_i|^2 \frac{\sigma^2_x}{\sigma^2_n} \right) \right),
\end{equation}
i.e., the capacity with perfect receiver CSI \cite{shannon1948}. Notice also that $\sigma^2_i$, $\bunderline{\mu}'_i$ and $|x_{i-1}|^2$ are random variables, whose joint distribution that can be calculated recursively, thus making it a computationally efficient method to find a lower bound on the capacity.

\subsection*{Proof:}

The chain rule of mutual information is used to lower bound the right-hand side (RHS) of (\ref{m10}):
\begin{IEEEeqnarray}{rcl}
\frac{1}{N} \mathcal{I}(\boldsymbol{x}_0^{N-1};\boldsymbol{y}_0^{N-1}) & \, = \, & \frac{1}{N} \sum_{i=0}^{N-1} \mathcal{I}(x_i;\boldsymbol{y}_0^{N-1} | \boldsymbol{x}_0^{i-1}), \nonumber \\
& \, = \, & \frac{1}{N} \sum_{i=0}^{N-1} \sum_{j=0}^{N-1} \mathcal{I}(x_i;y_j | \boldsymbol{x}_0^{i-1}, \boldsymbol{y}_0^{j-1}), \nonumber \\
\label{m110}
& \, \geq \, & \frac{1}{N} \sum_{i=0}^{N-1} \mathcal{I}(x_i;y_i | \boldsymbol{x}_0^{i-1}, \boldsymbol{y}_0^{i-1}),
\end{IEEEeqnarray}
where $\mathcal{I}(.;.)$ is mutual information.

\subsection*{Lemma 3}
\label{l1}

For the channel defined in (\ref{ofdmeq20}), the mutual information can be evaluated thus:
\begin{equation}
\label{m130}
\mathcal{I}(x_i;y_i | \boldsymbol{x}_0^{i-1}, \boldsymbol{y}_0^{i-1}) =  \mathcal{I}(x_i;y_i | \hat{z}_i),
\end{equation}
where:
\begin{equation}
\label{m140}
\hat{z}_i = P(z_i | \boldsymbol{x}_0^{i-1}, \boldsymbol{y}_0^{i-1}).
\end{equation}

\subsection*{Proof:}

In the appendix.

\subsection*{Lemma 4:}
For the channel defined in (\ref{ofdmeq20})--(\ref{ofdmeq60}), the conditional distribution of the frequency response, $z_i$, given all previous realisations of the input, $x_i$, and output, $y_i$ is a Gaussian distribution:
\begin{equation}
\label{m150}
(\bunderline{z}_i | \boldsymbol{\bunderline{x}}_{0}^{i-1}, \boldsymbol{\bunderline{y}}_{0}^{i-1}) \sim \mathcal{N}(\bunderline{z}_i ; \bunderline{\mu}_i,\Sigma_i-\Sigma_\epsilon),
\end{equation}
where :
\begin{IEEEeqnarray}{rcl}
\label{m160}
\Sigma_i & \, = \, & \begin{bmatrix} \sigma_i^2 & 0 \\ 0 & \sigma_i^2 \end{bmatrix}, \\
\label{m170}
\sigma^2_i & \, = \, & \Bigg\{ \begin{array}{ll} \sigma_z^2 & \textnormal{if} \,\,i=0, \\ \begin{array}{ll} (1-|a|^2)\sigma^2_z \\ +|a|^2(\sigma_{i-1}^{-2}+|x_{i-1}|^{2} \sigma_{n}^{-2})^{-1} \end{array} & \textnormal{if} \,\,i \neq 0, \end{array} \\
\label{m180}
& \, \leq \, & \sigma_z^2, \\
\label{m190}
\bunderline{\mu}_i & \sim & \mathcal{N}(\mu_i ; \ubar{0}, \Sigma_z - (\Sigma_i-\Sigma_\epsilon)) 
\end{IEEEeqnarray}
and $\Sigma_\epsilon$ is some Positive Definite Symmetric (PDS) matrix or zero.

\subsection*{Proof:}

In the appendix.

\subsection*{Proof of Theorem~2 (continued):}

Using Lemma~3 to consider only the $i^\textnormal{th}$ frequency response, $z_i$, which is itself a random variable, from (\ref{ofdmeq20}):
\begin{equation}
\label{theq10}
\bunderline{y}_i = \bunderline{Z}_i \bunderline{x}_i + \bunderline{n}_i,
\end{equation}
and as shown in Lemma~4, $z_i$ is a circularly symmetric Gaussian random variable, thus decomposing $z_i$ such that:
\begin{equation}
\label{theq20}
\bunderline{z}'_i \sim \mathcal{N} \left(\bunderline{z}'_i; \ubar{0}, \Sigma_i \right),
\end{equation}
it follows that:
\begin{equation}
\label{theq30}
\bunderline{y}_i = \bunderline{M}_i \bunderline{x}_i + \bunderline{Z}'_i \bunderline{x}_i + \bunderline{n}_i,
\end{equation}
where $M$ is the capitalised version of $\mu$, i.e., for the purposes of representing complex multiplication as a matrix operation. Further decomposing $\mu$, such that:
\begin{IEEEeqnarray}{rcl}
\label{theq40}
\bunderline{\mu}_i' & \, \sim \, & \mathcal{N}(\bunderline{\mu}_i'; \ubar{0}, \Sigma_z - \Sigma_i), \\
\label{theq50}
\bunderline{\mu}_i'' & \, \sim \, & \mathcal{N}(\bunderline{\mu}_i'';\ubar{0}, \Sigma_\epsilon),
\end{IEEEeqnarray}
(\ref{theq30}) can be expressed:
\begin{equation}
\label{theq60}
\bunderline{y}_i = \bunderline{M}_i' \bunderline{x}_i + \bunderline{M}_i''  \bunderline{x}_i + \bunderline{Z}'_i  \bunderline{x}_i + \bunderline{n}_i.
\end{equation}

Consider the mutual information:
\begin{equation}
\label{m240}
\mathcal{I}(x_i;y_i | \hat{z_i}) =  \mathcal{H}(y_i | \hat{z_i}) - \mathcal{H}(y_i | x_i, \hat{z_i}).
\end{equation}

Now consider the first term of the RHS of (\ref{m240}):
\begin{IEEEeqnarray}{rcl}
\mathcal{H}(y_i | \hat{z_i}) & \, \geq \, & \mathcal{H}(y_i | \hat{z_i}, z_i'), \nonumber \\
& \, = \, & \log_2 (2 \pi e |\bunderline{M}'_i \Sigma_x (\bunderline{M}'_i)^T \nonumber \\
& & \, \, \, \, \, \, 
+\bunderline{M}''_i \Sigma_x (\bunderline{M}''_i)^T +\bunderline{Z}'_i \Sigma_x (\bunderline{Z}'_i)^T +\Sigma_n|^{1/2}) \nonumber \\
\label{m250}
& \, \geq \, & \log_2 (2 \pi e ||\mu_i|^2 \Sigma_x  + \Sigma_n|^{1/2}),
\end{IEEEeqnarray}
and also consider the second term of the RHS of (\ref{m240}):
\begin{IEEEeqnarray}{rcl}
\mathcal{H}(y_i | x_i, \hat{z_i}) & = & \log_2 (2 \pi e |\bunderline{X}_i (\Sigma_i-\Sigma_\epsilon) \bunderline{X}^T_i +\Sigma_n|^{1/2}), \nonumber \\
\label{m259}
& = & \log_2 \left(2 \pi e ||x_i|^2 (\Sigma_i - \Sigma_\epsilon) +\Sigma_n|^{1/2}\right), \\
\label{m260}
& \leq & \log_2 (2 \pi e ||x_i|^2\Sigma_i  +\Sigma_n|^{1/2}), 
\end{IEEEeqnarray}
where (\ref{m260}) is derived from (\ref{m259}) by noticing that all terms within the determinant in (\ref{m259}) are proportional to the identity, except $\Sigma'_\epsilon$ which is PDS (as $\Sigma_i-\Sigma_\epsilon$ is a covariance matrix). Therefore applying Lemma~7, as detailed in the appendix, yields this result.

Substituting (\ref{m250}) and (\ref{m260}) into (\ref{m240}):
\begin{IEEEeqnarray}{rcl}
\mathcal{I}(x_i;y_i | \hat{z_i}) & \, = \, & \mathcal{H}(y_i | \hat{z_i}) -  \mathcal{H}(y_i | x_i, \hat{z_i}) , \nonumber\\
& \, \geq \, & \log_2 (2 \pi e ||\mu'_i|^2 \Sigma_x  +\Sigma_n|^{1/2})  \nonumber \\
& & \, \, \, 
- \log_2 (2 \pi e ||x_i|^2 \Sigma_i +\Sigma_n|^{1/2}) \nonumber \\
& \, = \, & \log_2 \left( \frac{ |\mu'_i|^2 \sigma_x^2 + \sigma_n^2}{|x_i|^2 \sigma_i^2+ \sigma_n^2} \right), \nonumber \\
\label{m270}
& \, = \, & \mathcal{I}_i,
\end{IEEEeqnarray}
which proves Theorem 2.

\subsection*{Corollary 8:}

There exists a lower bound, L$_2$, on the achievable rate:

\begin{equation}
\label{1cor10}
\mathrm{R} \geq \textnormal{L}_2 =  \frac{1}{N} \sum_{i=0}^{N-1} \mathcal{I}'_i \, \, \textnormal{bit}\, \textnormal{s}^{-1}\, \textnormal{Hz}^{-1},
\end{equation}
where:
\begin{equation}
\label{1cor20}
\mathcal{I}'_i = \mathbb{E} \left( \max \left(\log_2 \left(\frac{|\mu'_i|^2 \sigma_x^2 + \sigma_n^2}{|x_i|^2 \sigma_i^2 +\sigma_n^2} \right) , 0  \right) \right).
\end{equation}

This corollary simply allows $\mathcal{I}_i$ to be replaced with zero if it negative. This is useful for computation, and slightly tightens the bound. 

\subsection*{Proof:}

\begin{equation}
\label{1cor30}
\mathcal{H}(y_i | \hat{z_i}) -  \mathcal{H}(y_i | x_i, \hat{z_i}) \geq 0,
\end{equation}
substituting (\ref{1cor30}) into (\ref{m270}):
\begin{IEEEeqnarray}{rcl}
\mathcal{I}(x_i;y_i | \hat{z_i})  & \, \geq \, &  \mathbb{E} \left( \max \left(\log_2 \left(\frac{|\mu'_i|^2 \sigma_x^2 + \sigma_n^2}{|x_i|^2 \sigma_i^2 +\sigma_n^2} \right) , 0 \right) \right), \nonumber \\
\label{1cor40}
& \, = \, & \mathcal{I}'_i,
\end{IEEEeqnarray}
which proves Corollary 8.

\subsection*{Corollary 9:}

There exist lower bounds $\mathrm{L}_{1\textnormal{A}}$ and $\mathrm{L}_{2\textnormal{A}}$ such that:
\begin{IEEEeqnarray}{rcccl}
\label{m290a}
\mathrm{R} \geq \mathrm{L}_1 & \, \geq \, & \mathrm{L}_{1\textnormal{A}} & \, = \, & \frac{1}{N} \left( \sum_{i=0}^{N''-1} \mathcal{I}_i + (N-N'')\mathcal{I}_{N''} \right), \\
\label{m290b}
\mathrm{R} \geq \mathrm{L}_2 & \, \geq \, & \mathrm{L}_{2\textnormal{A}} & \, = \, & \frac{1}{N} \left( \sum_{i=0}^{N''-1} \mathcal{I}'_i + (N-N'')\mathcal{I}'_{N''} \right), 
\end{IEEEeqnarray}
where:
\begin{equation}
\label{m300}
0 < N'' \leq N.
\end{equation}

This is useful as the lower bounds L$_1$ and L$_2$ can themselves be lower bounded by L$_{1\textnormal{A}}$ and L$_{2\textnormal{A}}$ respectively, by noticing that the sequence $\mathcal{I}(x_i;y_i | \hat{z_i})$ is monotonically non-decreasing, and therefore a lower bound on the $i^{\textnormal{th}}$ term is automatically a lower bound on the $(i+j)^{\textnormal{th}}$ term, $j \geq 0$. This means computation of the mutual information could be halted when a sufficiently tight lower bound has been achieved.

\subsection*{Proof:}

For $j>0$, consider (\ref{m270}), and using Lemma~3:

\begin{IEEEeqnarray}{rcl}
\mathcal{I}(x_i;y_i | \hat{z}_i) & \, = \,  &  \mathcal{H}(x_i | \hat{z_i}) -  \mathcal{H}(x_i | y_i, \hat{z_i}) , \nonumber\\
\label{m310}
& \, = \, & \mathcal{H}(x_{i}) - \mathcal{H}(x_{i} | y_{i}, \boldsymbol{x}_0^{i-1}, \boldsymbol{y}_0^{i-1}),
\end{IEEEeqnarray}
where the conditioning in the first term has been dropped because $x_i$ is an IID random variable, and there is no feedback in the channel. Likewise:
\begin{IEEEeqnarray}{rcl}
\mathcal{I}(x_{i+j};y_{i+j} | \hat{z}_{i+j}) & \, = \, &  \mathcal{H}(x_{i+j}) - \nonumber \\
\label{m320}
&& \, \, 
\mathcal{H}(x_{i+j} | y_{i+j}, \boldsymbol{x}_0^{i+j-1}, \boldsymbol{y}_0^{i+j-1}), 
\end{IEEEeqnarray}

Therefore:

\begin{IEEEeqnarray}{lcl}
\begin{array}{l} \mathcal{I}(x_{i+j};y_{i+j} | \hat{z}_{i+j}) \\ - \mathcal{I}(x_i;y_i | \hat{z_i}) \end{array} & \,   = \, & \begin{array}{l} \mathcal{H}(x_{i} | y_{i}, \boldsymbol{x}_0^{i-1}, \boldsymbol{y}_0^{i-1}) \\ - \mathcal{H}(x_{i+j} | y_{i+j}, \boldsymbol{x}_0^{i+j-1}, \boldsymbol{y}_0^{i+j-1}), \end{array} \nonumber \\
\label{m330}
& \, \geq \, & 0, 
\end{IEEEeqnarray}
because the two terms in the RHS differ only by conditioning, and conditioning reduces entropy. Noticing that exactly the same analysis can be applied to $\mathcal{I}'_i$, this is a sufficient condition to prove Corollary~9.

\section{Numerical example}
\label{ex}

As identified in Section~\ref{int}, in-vehicle channels are of particular interest, which have PDPs which decay exponentially with time \cite{us1, sair}. For such channels, the delay spread is infinite, and therefore the underspread property is only approximate. At sufficiently large SNR, the fact that this is only an approximation becomes significant, as identified by Durisi \textit{et al} \cite{durisi3}, and further supported Koch and Lapidoth \cite{kl} who show that, for a discrete exponentially decaying channel, the capacity is bounded in the Signal to Noise Ratio (SNR). It is therefore important not to over-generalise the applicability of our bound, and thus it is evaluated for a typical example application. A suitable example application is a Wireless Sensor Network operating using \textit{Zigbee} \cite{zigbee}. To evaluate the lower bound, it is necessary to find appropriate parameters to substitute into the expressions (\ref{m70}), (\ref{m80}), (\ref{m90}) and (\ref{m100}). These parameters derive from the fundamental parameters (i.e., the cavity time constant and system SNR) via the parameters required to model the channel as a block fading system (i.e., the block length, cyclic prefix length and the adjustment to the SNR to account for the block fading model).

\subsection{Parameters}
\label{paramsec}

The capacity of a channel with perfect receiver CSI can be expressed:
\begin{IEEEeqnarray}{rcl}
C & \, = \, & \mathbb{E} \left( \log_2 \left( 1 + |z|^2 \frac{\sigma_x^2}{\sigma_n^2} \right) \right), \nonumber \\
& \, = \, & \mathbb{E} \left( \log_2 \left( 1 + |z''|^2 \frac{\sigma_z^2 \sigma_x^2}{\sigma_n^2} \right) \right), \nonumber \\
\label{paramextra10}
& \, = \, &  \mathbb{E} \left( \log_2 \left( 1 + |z''|^2 \textnormal{SNR} \right) \right),
\end{IEEEeqnarray}
where, by definition:
\begin{IEEEeqnarray}{rcl}
\label{paramextra20}
\bunderline{z}'' & \, \sim \, & \mathcal{N} \left( \bunderline{z}'' ; \ubar{0} , \begin{bmatrix} 1 & 0 \\ 0 & 1 \end{bmatrix} \right), \\
\label{paramextra30}
\textnormal{SNR} & \,  = \, & \frac{\sigma^2_z \sigma^2_x}{\sigma^2_n}.
\end{IEEEeqnarray}
It can be shown that the specified 250~kbit/s for a single Zigbee channel (i.e., occupying a frequency band of width 5~MHz) can be achieved at an SNR of 0.0180 (for this analysis it is irrelevant that actual Zigbee systems would typically have a much higher SNR).\\

From our previous measurements \cite{us1}, the time constant, $\tau_c$, of the exponential decay in a typical vehicle cavity is 17.2~ns. Regarding the choice of block length, $T_B$, an appropriate criteria (i.e., for this example) is the time duration during which 0.99 of the energy is expected to remain undisturbed. Based on the measurements in \cite{us2}, this is equal to 0.0053~s at 2.45~GHz. These results can be used to show: $N=26500$ and $\Delta \omega = 189$~Hz. Regarding the choice of cyclic prefix length, $T_t$, and noting that this must correspond to an integer number of samples in the time domain (i.e., according to the Sampling Theorem), consider choosing just a single time sample duration to be the cyclic prefix. This is equal to 200~ns, and the energy which has thus been truncated, $\mathbb{E} \left( \mathrm{E}_\textnormal{trunc} \right)$, can thus be evaluated:
\begin{IEEEeqnarray}{rcl}
\mathbb{E} \left( \mathrm{E}_\textnormal{trunc} \right) & \, = \, & 1- \frac{1}{\tau_c} \int_0^{T_t} e^{-\frac{\tau}{\tau_c}} \, \mathrm{d} \tau, \nonumber \\
& \, = \, & e^{-\frac{T_t}{\tau_c}}, \nonumber \\
& \, = \, & 8.91 \times 10^{-6},
\end{IEEEeqnarray}
where $T_t = 200$~ns, note that the term $1/\tau_c$ is included for normalisation, such that the total energy is unity. To allow for the fact that only 0.99 of the energy actually does not vary during one time block (as opposed to it being completely invariant), and that $1-8.91 \times 10^{-6}$ of the energy is contained in the truncated part of the PDP, it is necessary to adjust the SNR. This is achieved by assuming that an infinite number of time blocks have preceded the current one, and treating the energy leaking into the current time block (i.e., from previous time blocks) as noise, and also treating the energy which varies within one time block as noise:
\begin{IEEEeqnarray}{rcl}
\textnormal{SNR}' & \, = \, & \frac{\textnormal{SNR} \times 0.99 \times (1-8.91 \times 10^{-6})}{1+ \textnormal{SNR}(1-0.99 \times (1-8.91 \times 10^{-6}))}, \nonumber \\
& \, = \, & \frac{0.018 \times 0.99 \times (1-8.91 \times 10^{-6})}{1+ 0.018(1- 0.99 \times (1-8.91 \times 10^{-6}))}, \nonumber \\
& \, = \, & 0.0178,
\end{IEEEeqnarray}
where SNR' is the adjusted SNR.\\

To evaluate the lower bound, it is necessary to find $a$, from (\ref{fr110}) it can be shown:
\begin{equation}
\label{aeq10}
a = \frac{1}{1+ j \Delta \omega \tau_c} \left(1- e^{-\tau_t((1/\tau_c)+j \Delta \omega)}\right) \left(1-e^{\tau_t/\tau} \right)^{-1},
\end{equation}
and noticing that as $\tau_t \to \infty$ (i.e., we do not truncate the impulse response):
\begin{IEEEeqnarray}{rcl}
a & = & \frac{1}{1+j \tau_c \Delta \omega}, \nonumber \\
\label{aeq20}
& = & \frac{1}{1+j \tau_c / T_B},
\end{IEEEeqnarray}
i.e., by substituting in $\Delta \omega = 1/ T_B$ as explained in Section~\ref{chmod2}. This demonstrates the intuitive property that as the coherence time gets large relative to the delay spread $a \to 1$ and thus there is negligible variation between successive frequency samples. Measured data demonstrates that (\ref{aeq20}) is valid for the in-vehicle channel (without truncating the impulse response).\\

Substituting $\tau_c$, $T_B$ and $\tau_t = T_t$ into (\ref{aeq10}) yields $|a|^2 = 1 - 1.03 \times 10^{-11}$. To determine appropriate values of $\sigma^2_z$, $\sigma^2_x$ and $\sigma^2_n$, the adjusted SNR is sufficient. As one parameter is used to determine three parameters, there is some choice and it necessary to establish whether the bounding method imposes any restrictions on this choice.

\subsection*{Proposition 10}

The bound in (\ref{m70}) relies only on the value of SNR as defined in (\ref{paramextra30}), and not the individual realisations of $\sigma^2_z$, $\sigma^2_x$ and $\sigma^2_n$.

\subsection*{Proof:}

Using the location scale property of the Gaussian distribution, let:
\begin{IEEEeqnarray}{rcl}
x_i' & \, = \, & \frac{1}{\sigma_x} x_i, \nonumber \\
\label{prop10eq20}
& \, \sim \, & \mathcal{N} \left( x_i'; \ubar{0}, \begin{bmatrix} 1 & 0 \\ 0 & 1 \end{bmatrix} \right), \\
(\sigma_i')^2 & \, = \, & \frac{1}{\sigma^2_z} \sigma^2_i, \nonumber \\
\label{prop10eq30}
& \, = \, & (1- |a|^2) + |a|^2\left((\sigma'_{i-1})^{-2}+  \frac{\sigma^2_z \sigma^2_x}{\sigma^2_n}  |x'_{i-1}|^{2} \right)^{-1} \\
\mu'''_i & \, = \, & \frac{1}{\sigma_z} \mu_i, \nonumber \\
\label{prop10eq40}
& \, \sim \, & \mathcal{N} \left( \mu_i'''; \ubar{0}, \begin{bmatrix} 1-(\sigma'_i)^2 & 0 \\ 0 & 1 - (\sigma'_i)^2 \end{bmatrix} \right).
\end{IEEEeqnarray}

Substituting the results into (\ref{m70}) yields:
\begin{IEEEeqnarray}{rcl}
\mathcal{I}_i & \, = \, & \mathbb{E} \left( \log_2 \left(\frac{\sigma_z^2|\mu'''_i|^2 \sigma_x^2 + \sigma_n^2}{\sigma_x^2|x'_i|^2 \sigma_z^2 (\sigma'_i)^2 +\sigma_n^2} \right) \right), \nonumber \\
\label{prop10eq50}
& \, = \, & \mathbb{E} \left( \log_2 \left(\frac{ \frac{\sigma^2_z \sigma^2_x}{\sigma^2_n}  |\mu'''_i|^2  + 1}{ \frac{\sigma^2_z \sigma^2_x}{\sigma^2_n} |x'_i|^2  (\sigma'_i)^2 +1} \right) \right). 
\end{IEEEeqnarray}

Therefore (\ref{prop10eq30}), (\ref{prop10eq40}) and (\ref{prop10eq50}) show that the bound only relies on $\sigma^2_z \sigma^2_x / \sigma^2_n$, thus proving Proposition~10.\\

Given that the choice of $\sigma^2_z$, $\sigma^2_x$ and $\sigma^2_n$ is arbitrary, so long as they combine to form the correct SNR,  $\sigma^2_z=0.5$, $\sigma^2_x=0.0356$ and $\sigma^2_n=1$ are chosen. A summary of all the parameters is given in Table~\ref{capt1}.

\begin{table}[t]
\renewcommand{\arraystretch}{1.3}
\caption{Summary of parameters.}
\label{capt1}
\centering
\begin{tabular}{ |l | l | l |   }
\hline
\textbf{Type} & \textbf{Name} & \textbf{Value}  \\
\hline
\hline
Fundamental & $\tau_c$ & $1.7 \times 10^{-8}$~s  \\
\cline{2-3}
& $W$ & $5 \times 10^{6}$~Hz \\
\cline{2-3}
& SNR & $1.80 \times 10^{-2}$ \\
\hline
\hline
Block fading \& OFDM & $T_B$ & $5.30 \times 10^{-3}$~s \\
\cline{2-3}
& $T_t$ & $2 \times 10^{-7}$~s\\
\cline{2-3}
& $N$ & 26500 \\
\cline{2-3}
& $\Delta \omega$ & $1.89 \times 10^{2}$~Hz \\
\cline{2-3}
& SNR' & $1.78 \times 10^{-2}$ \\
\hline
\hline
Lower bound & $|a|^2$ & $1 - 1.03 \times 10^{-11}$ \\
\cline{2-3}
& $\sigma^2_z$ & $5 \times 10^{-1}$ \\
\cline{2-3}
& $\sigma^2_x$ & $3.56 \times 10^{-2}$ \\
\cline{2-3}
& $\sigma^2_n$ & 1 \\
\hline
\end{tabular}
\end{table}

\subsection{Results}

The lower bound L$_2$, i.e., from (\ref{1cor10}), has been found for the parameters specified in Section~\ref{paramsec}. It should, however, be noted that to rigorously lower bound the channel, it is necessary to take into account the fact that no information is transferred during the cyclic prefix:
\begin{IEEEeqnarray}{rcl}
C & \, \geq \, & \mathrm{L}_{2\textnormal{B}} \nonumber \\
\label{prefixbeq10}
& \, = \, & \mathrm{L}_2 \times \frac{T_B}{T_B + T_t},
\end{IEEEeqnarray}
where $\mathrm{L}_{2\textnormal{B}}$ is the lower bound.\\

\begin{figure}[!t]
\centering
\includegraphics[width=3in]{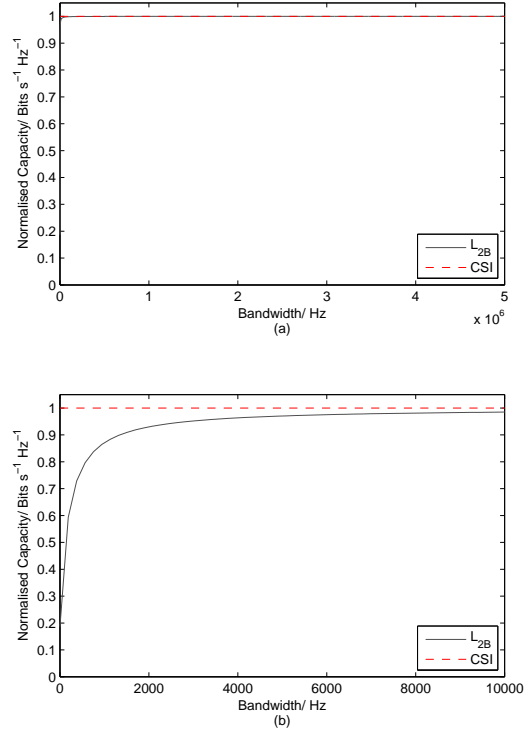}
\caption{Capacity lower bound (a) full; (b) detailed.}
\label{ch5f3}
\end{figure}

The result of the bounding process is shown in Fig.~\ref{ch5f3}(a). Also shown, in Fig.~\ref{ch5f3}(b), is a detailed close-up of the bound at low values of bandwidth. As seen, the lower bound is tight, achieving 0.9999 of the capacity with perfect receiver CSI.

\subsection{Comparison with previous lower bounds}
\label{compare}

Comparing our lower bound, $\mathrm{L}_{2 \textnormal{B}}$, to previous lower bounds is not straightforward, given that the channel definition is not identical. Nonetheless, it can be seen that our channel roughly corresponds to the channel in [\cite{durisi3}~Fig.~2.10(b)] at an SNR of -17~dB. In this figure, the channel spread is $10^{-6}$, and the amount of energy not compactly supported within this spread (denoted $\epsilon$ in \cite{durisi3}) is also $10^{-6}$. We can conclude that our channel is actually a worse case than this, as our spread is of the order $10^{-5}$ \cite{us2} and our choice of coherence time is such that $\epsilon \approx 10^{-2}$. Our bound of 0.9999 times the capacity with perfect receiver CSI clearly exceeds that derived by Durisi \textit{et al} [\cite{durisi3}~Fig.~2.10(b)] which is approximately 0.85 times the capacity with perfect receiver CSI.\\

It is, however, important to note that Durisi \textit{et al} \cite{durisi3} have included a constraint on the peak power in time and frequency, which is 10 times the average power. Dispensing with this constraint may allow this bound to be tightened.

%The constraint on the peak power in time can be dealt with by simply disqualifying and time blocks where the constraint is violated. The total power will be the sum of $N$ Gaussian random variables, and thus according to the central limit theorem will itself be a Gaussian random variable. The power of this Gaussian random variable will exceed 10 times its average value less than 0.0007 of the time (i.e., when the Gaussian random variable does not lie within 3.3 standard deviations of zero). The constraint in frequency is somewhat harder to deal with, as the time duration over which the channel is sampled is unclear. If this is infinite, then the power usage will be flat across frequency, however in the worst case, where the sampling duration is equal to or shorter than a coherence time block, then it is necessary to disqualify some frequency samples in the same manner. Again, only less than 0.0027 of the time will the power a frequency sample exceed 10 times its average value.\\

%This analysis allows a further lower bound, $\mathrm{L}_{\textnormal{peak}}$ to be evaluated:
%\begin{IEEEeqnarray}{rcl}
%\mathrm{L}_{\textnormal{peak}} & = & \mathrm{L}_{2 \textnormal{B}} \times (1-0.0027)^2 , \nonumber \\
%\label{compaeq10}
%& = & \mathrm{L}_{2 \textnormal{B}} \times 0.995,
%\end{IEEEeqnarray}
%which means that $\mathrm{L}_{\textnormal{peak}} = 0.995 C_{\textnormal{CSI}}$ for this channel (where $C_{\textnormal{CSI}}$ is the channel capacity with perfect receiver CSI). This therefore exceeds the values of approximately $0.85 C_{\textnormal{CSI}}$ from [\cite{durisi3}~Fig.~2.10(b)].

\section{Conclusions}
\label{conc}

Most actual real-world channels are highly underspread, that is that they remain virtually unchanged for a time duration much greater than their delay spread. Early analysis showed that, in the infinite bandwidth limit, the noncoherent capacity of an underspread channel with AWGN tends to that of the same channel with perfect CSI at the receiver. Since this early analysis, the majority of the research in the field of noncoherent capacity of underspread channels has focussed on bounding the capacity for actual wireless communication situations, and this paper provides a lower bound on the noncoherent capacity of highly underspread channels which complements the existing work.\\

Specifically, the lower bound proposed in this paper assumes only that the PDP and the coherence time of the channel are known, which is a more general starting point than that assumed previously. Furthermore, the bound is intuitive, mathematically elegant, computationally simple and allows easy computation of the minimum bandwidth required to achieve a specified fraction of the capacity with perfect CSI at the receiver. A numerical example has been included which demonstrates that our bound is tighter, at least in some situations, than those proposed previously.\\

%Our lower bound also allows a peak constraint to be placed on the transmit power in time and frequency, however this is not a major feature of our work, and only works well if the peak constraint is relatively large.\\

Finally, we note that treating the channel as underspread is only an approximation, as many actual channels will, in general, have infinite time-duration impulse responses. Whilst it is fair to say that, at typical values of SNR, the bounds proposed in this paper and the previous literature are very useful, finding a general expression for noncoherent capacity (for channels which may be overspread) is an interesting open problem, and one which will allow the behaviour of the noncoherent capacity of various channels to be characterised as the SNR tends to infinity.

\appendices
\section{Proof of Lemmas}
\subsection*{Lemma 3}
\label{l1a}

For the channel defined in (\ref{ofdmeq20}), the mutual information can be evaluated thus:
\begin{equation}
\label{m130a}
\mathcal{I}(x_i;y_i | \boldsymbol{x}_0^{i-1}, \boldsymbol{y}_0^{i-1}) =  \mathcal{I}(x_i;y_i | \hat{z}_i),
\end{equation}
where:
\begin{equation}
\label{m140a}
\hat{z}_i = P(z_i | \boldsymbol{x}_0^{i-1}, \boldsymbol{y}_0^{i-1}).
\end{equation}

\subsection*{Proof:}

\begin{IEEEeqnarray}{rcl}
\mathcal{I}(x_i;y_i | \boldsymbol{x}_0^{i-1}, \boldsymbol{y}_0^{i-1}) & \, = \, & \mathcal{H}(x_i| \boldsymbol{x}_0^{i-1}, \boldsymbol{y}_0^{i-1}) \nonumber \\
& & \, \, \, 
-\mathcal{H}(x_i | y_i, \boldsymbol{x}_0^{i-1}, \boldsymbol{y}_0^{i-1}), \nonumber \\
\label{1app5}
& \, = \, & \mathcal{H}(x_i) - \mathcal{H}(x_i | y_i, \boldsymbol{x}_0^{i-1}, \boldsymbol{y}_0^{i-1}), 
\end{IEEEeqnarray}
where the conditioning in the first term of the RHS is dropped as $x_i$ are IID random variables, and there is not feedback in the channel. Regarding the second term of the RHS of (\ref{1app5}), consider:
\begin{IEEEeqnarray}{rcl}
P(x_i | y_i, \boldsymbol{x}_0^{i-1}, \boldsymbol{y}_0^{i-1}) & \, = \, & \int_{z_i} P(x_i | z_i, y_i, \boldsymbol{x}_0^{i-1}, \boldsymbol{y}_0^{i-1}) \nonumber \\
& & \, \, \, \, \, 
 P(z_i | y_i, \boldsymbol{x}_0^{i-1}, \boldsymbol{y}_0^{i-1}) \, \mathrm{d}z_i, \nonumber \\
& \, = \, & \int_{z_i} P(x_i | z_i, y_i) \nonumber \\
\label{1app6}
& & \, \, \, \, \, \, 
P(z_i | y_i, \boldsymbol{x}_0^{i-1}, \boldsymbol{y}_0^{i-1}) \, \mathrm{d}z_i, 
\end{IEEEeqnarray}
where the conditioning in the first term of the integral has been dropped, because $x_i$ is conditionally independent of $(\boldsymbol{x}_0^{i-1}, \boldsymbol{y}_0^{i-1})$ given $(z_i, y_i)$. Notice also that, for the second term on the RHS of (\ref{1app6}), $y_i$ and $(\boldsymbol{x}_0^{i-1}, \boldsymbol{y}_0^{i-1})$ are conditionally independent given $z_i$, thus:
\begin{IEEEeqnarray}{rcl}
P(z_i | y_i, \boldsymbol{x}_0^{i-1}, \boldsymbol{y}_0^{i-1}) &  =  & \frac{P(y_i, \boldsymbol{x}_0^{i-1}, \boldsymbol{y}_0^{i-1}| z_i)P(z_i)}{P( y_i, \boldsymbol{x}_0^{i-1}, \boldsymbol{y}_0^{i-1})} \nonumber \\
& = & \frac{P(y_i | z_i) P(\boldsymbol{x}_0^{i-1}, \boldsymbol{y}_0^{i-1}| z_i) P(z_i)}{P( y_i, \boldsymbol{x}_0^{i-1}, \boldsymbol{y}_0^{i-1})} \nonumber \\
& = & \frac{P(z_i |y_i) P(y_i)}{P(z_i)} \nonumber \\
& & \, \,
\times \frac{P(z_i |\boldsymbol{x}_0^{i-1}, \boldsymbol{y}_0^{i-1}) P(\boldsymbol{x}_0^{i-1}, \boldsymbol{y}_0^{i-1})}{P(z_i)} \nonumber \\
& & \, \,
\times \frac{P(z_i)}{P( y_i, \boldsymbol{x}_0^{i-1}, \boldsymbol{y}_0^{i-1})} \nonumber \\
& \propto & \frac{P(z_i |y_i) P(z_i |\boldsymbol{x}_0^{i-1}, \boldsymbol{y}_0^{i-1})}{P(z_i)} \nonumber \\
& = & \frac{P(z_i |y_i) \hat{z}_i}{P(z_i)} \nonumber \\
\label{1app7}
& = & P(z_i | y_i, \hat{z}_i).
\end{IEEEeqnarray}

Therefore, substituting (\ref{1app7}) into (\ref{1app6}): 
\begin{IEEEeqnarray}{rrcl}
& P(x_i | y_i, \boldsymbol{x}_0^{i-1}, \boldsymbol{y}_0^{i-1}) & \, = \, & P(x_i | y_i, \hat{z}_i ), \nonumber \\
\label{1app8}
\implies \, & \mathcal{H}(x_i | y_i, \boldsymbol{x}_0^{i-1}, \boldsymbol{y}_0^{i-1}) & \, = \, & \mathcal{H}(x_i | y_i, \hat{z}_i ),
\end{IEEEeqnarray}
substituting (\ref{1app8}) into (\ref{1app5})
\begin{IEEEeqnarray}{rcl}
\mathcal{I}(x_i;y_i | \boldsymbol{x}_0^{i-1}, \boldsymbol{y}_0^{i-1}) & \, = \, & \mathcal{H}(x_i) - \mathcal{H}(x_i | y_i, \hat{z_i} ),  \nonumber \\
\label{1app10}
& \, = \, & \mathcal{I}(x_i ; y_i | \hat{z_i}),
\end{IEEEeqnarray}
which proves Lemma 3.

\subsection*{Lemma 4:}
For the channel defined in (\ref{ofdmeq20}), the conditional distribution of the frequency response, $z_i$, given all previous realisations of the input, $x_i$, and output, $y_i$ is a Gaussian distribution:
\begin{equation}
\label{m150a}
(\bunderline{z}_i | \boldsymbol{\bunderline{x}}_{0}^{i-1}, \boldsymbol{\bunderline{y}}_{0}^{i-1}) \sim \mathcal{N}(\bunderline{z}_i ; \bunderline{\mu}_i,\Sigma_i-\Sigma_\epsilon),
\end{equation}
where :
\begin{IEEEeqnarray}{rcl}
\label{m160a}
\Sigma_i & \, = \, & \begin{bmatrix} \sigma_i^2 & 0 \\ 0 & \sigma_i^2 \end{bmatrix}, \\
\label{m170a}
\sigma^2_i & \, = \, & \Bigg\{ \begin{array}{ll} \sigma_z^2 & \textnormal{if} \,\,i=0, \\ \begin{array}{l} (1-|a|^2)\sigma^2_z \\ +|a|^2(\sigma_{i-1}^{-2}+|x_{i-1}|^{2} \sigma_{n}^{-2})^{-1} \end{array} & \textnormal{if} \,\,i \neq 0, \end{array} \\
\label{m180a}
& \, \leq \, & \sigma_z^2, \\
\label{m190a}
\bunderline{\mu}_i & \sim & \mathcal{N}(\mu_i ; \ubar{0}, \Sigma_z - (\Sigma_i-\Sigma_\epsilon)) 
\end{IEEEeqnarray}
and $\Sigma_\epsilon$ is some Positive Definite Symmetric (PDS) matrix or zero.

\subsection*{Proof:}

Lemma~4 is proven using mathematical induction, for $i=0$:
\begin{equation}
\label{2app10}
\bunderline{z}_0 \sim  \mathcal{N}(\bunderline{z}_0 ; \ubar{0}, \Sigma_z),
\end{equation}
which is true by definition, as there are no previous values of $x_i$ and $y_i$ upon which $z_0$ is conditioned.\\

%for $\bunderline{z}_0$: $\sigma_{i-1}^2=\sigma_z$, and $x_{i-1} = 0$, therefore the Lemma is true for $i=0$.\\

Next, it is shown that if Lemma~4 is true for $\underline{z}_{i-1}$ then it is also true for $\underline{z}_{i}$
\begin{IEEEeqnarray}{rcl}
P(z_i|\boldsymbol{x}_0^{i-1},\boldsymbol{y}_0^{i-1}) &  = &  \int_{z_{i-1}}P(z_i|z_{i-1},\boldsymbol{x}_0^{i-1},\boldsymbol{y}_0^{i-1}) \nonumber \\
\label{2app20}
& & \, \, \, \, \, 
 P(z_{i-1}|\boldsymbol{x}_0^{i-1},\boldsymbol{y}_0^{i-1})\, \mathrm{d}z_{i-1}.
\end{IEEEeqnarray}
Consider the first term in the integrand in (\ref{2app20})
\begin{equation}
\label{2app21}
P(z_i|z_{i-1},\boldsymbol{x}_0^{i-1},\boldsymbol{y}_0^{i-1}) = P(z_i|z_{i-1},\boldsymbol{x}_0^{i-2},\boldsymbol{y}_0^{i-2}),
\end{equation}
and it is known from Proposition~1 that:
\begin{equation}
\label{2app30}
(\bunderline{z}_i|\bunderline{z}_{i-1}) \sim \mathcal{N}(\bunderline{z}_{i-1} ; \bunderline{A}  \bunderline{z}_{i-1}, \Sigma_{a}).
\end{equation}
Consider the multivariate Gaussian:
\begin{IEEEeqnarray}{rcl}
\begin{bmatrix} \bunderline{z}_i & | & \bunderline{z}_{i-1}, \boldsymbol{x}_0^{i-2} \\ \boldsymbol{y}_0^{i-2} & | & \bunderline{z}_{i-1}, \boldsymbol{x}_0^{i-2} \end{bmatrix} & \, = \, & \begin{bmatrix} \bunderline{z}_i & | & \bunderline{z}_{i-1} \\ \boldsymbol{y}_0^{i-2} & | & \bunderline{z}_{i-1}, \boldsymbol{x}_0^{i-2} \end{bmatrix} \nonumber \\
& \, \sim \, & \mathcal{N} \Bigg( \begin{bmatrix}\bunderline{z}_{i} \\ \boldsymbol{y}_0^{i-2} \end{bmatrix} ; \nonumber \\
\label{2app31}
& & \, \, \, \, \, 
\begin{bmatrix} \bunderline{A}  \bunderline{z}_{-1} \\ \boldsymbol{\bunderline{\alpha}} \end{bmatrix} , \begin{bmatrix} \Sigma_{a} & \beta \\ \beta^T & \delta \end{bmatrix} \Bigg),
\end{IEEEeqnarray}
where the values of $\boldsymbol{\bunderline{\alpha}}$, $\beta$ and $\delta$ are unimportant for this analysis. Therefore:
\begin{equation}
P(\bunderline{z}_i|\bunderline{z}_{i-1},\boldsymbol{x}_0^{i-1},\boldsymbol{y}_0^{i-1}) =  \mathcal{N}(\bunderline{z}_i ; \bunderline{A} \bunderline{z}_{i-1} +\bunderline{\mu}_\epsilon, \Sigma_{a} - \Sigma'_\epsilon), \nonumber
\end{equation}
which implies that:
\begin{IEEEeqnarray}{rcl}
P(\bunderline{A}^{-1}\bunderline{z}_i|\bunderline{z}_{i-1},\boldsymbol{x}_0^{i-1},\boldsymbol{y}_0^{i-1}) & \, = \, & \mathcal{N}(\bunderline{A}^{-1}\bunderline{z}_i ;  \bunderline{z}_{i-1} +\bunderline{A}^{-1}\bunderline{\mu}_\epsilon, \nonumber \\
& & \, \, \, \, \, \, \, \, \, \, \,  \bunderline{A}^{-1}\Sigma_{a}\bunderline{A}^{-T} \nonumber \\
\label{2app35}
& & \, \, \, \, \, \, \, \, \, \, \, 
- \bunderline{A}^{-1}\Sigma'_\epsilon\bunderline{A}^{-T}), 
\end{IEEEeqnarray}
where the value of $\bunderline{\mu}_\epsilon$ is unimportant for this analysis, as is $\Sigma'_\epsilon=\beta^T \delta^{-1} \beta$ which is a PDS matrix (i.e., because $\delta$ is a covariance matrix), or zero if the underlying process is actually a Markov process.\\

Consider the second term of the integrand in (\ref{2app20}), and notice that it can be split into two conditionally independent terms:
\begin{IEEEeqnarray}{rcl}
\label{2app40}
(\bunderline{z}_{i-1} | \bunderline{x}_{i-1}, \bunderline{y}_{i-1}) & \sim & \mathcal{N} \left(\bunderline{z}_{i-1}; \bunderline{X}_{i-1}^{-1}\bunderline{y}_{i-1}, |x_{i-1}|^2 \Sigma_n \right), \\
\label{2app50}
(\bunderline{z}_{i-1} | \boldsymbol{\bunderline{x}}_0^{i-2}, \boldsymbol{\bunderline{y}}_0^{i-2}) & \sim & \mathcal{N} \left(\bunderline{z}_{i-1};\bunderline{\mu}_{i-1}, \Sigma_{i-1} - \Sigma''_\epsilon \right),
\end{IEEEeqnarray}
i.e., from the definition of Lemma~4 in (\ref{m150}). This expression is valid for $i=1$, as $\boldsymbol{\bunderline{x}}_0^{i-2}, \boldsymbol{\bunderline{y}}_0^{i-2}$ consists of no elements, and thus it represents the unconditional distribution of $\bunderline{z}_{0}$, which is valid by the definition in (\ref{m170}), i.e., with $\Sigma''_\epsilon =0$.\\

The conditional independence allows (\ref{2app40}) and (\ref{2app50}) to be fused together as a Kalman filter \cite{kalman}, i.e., in which the pair $(x_{i-1}, y_{i-1})$ forms a measurement, and there exists some prior estimate of the state $P(z_{i-1} | \boldsymbol{x}_0^{i-2}, \boldsymbol{y}_0^{i-2})$. This leads to:
\begin{equation}
\label{2app60}
(\bunderline{z}_{i-1}|\boldsymbol{\bunderline{x}}_0^{i-1},\boldsymbol{\bunderline{y}}_0^{i-1}) \sim\mathcal{N}(\bunderline{z}_{i-1}; \bunderline{\mu}_{\alpha}, \Sigma_{\alpha}),
\end{equation}
where the value of $\bunderline{\mu}_\alpha$ is unimportant for this analysis, and:
\begin{IEEEeqnarray}{rcl}
\label{2app68}
\Sigma_{\alpha} & = & \left( |x_{i-1}|^2 \Sigma_n^{-1} + (\Sigma_{i-1} - \Sigma''_\epsilon)^{-1} \right)^{-1}, \\
\label{2app69}
& = &  \left( |x_{i-1}|^2 \Sigma_n^{-1} + \Sigma_{i-1}^{-1} + (\Sigma'''_\epsilon)^{-1} \right)^{-1}, \\
\label{2app70}
& = & \left( |x_{i-1}|^2 \Sigma_n^{-1} + \Sigma_{i-1}^{-1}\right)^{-1} - \Sigma''''_\epsilon ,
\end{IEEEeqnarray}
where $\Sigma'''_\epsilon$ and $\Sigma''''_\epsilon$ are PDS matrices. Lemma~5 is applied to $(\Sigma_{i-1} - \Sigma''_\epsilon)^{-1}$ in (\ref{2app68}), noticing that $\Sigma_{i-1}$ is proportional to the identity, to derive (\ref{2app69}). Lemma~6 is applied to the RHS of (\ref{2app69}), noticing that $(|x_{i-1}|^2 \Sigma_n^{-1} + \Sigma_{i-1}^{-1})$ is proportional to the identity, to derive (\ref{2app70}). The Lemmas are stated and proved in below in this appendix.\\

Substituting (\ref{2app35}) and (\ref{2app60}) into (\ref{2app20}), and performing the resulting convolution yields:
\begin{IEEEeqnarray}{rcl}
P(z_i|\boldsymbol{x}_0^{i-1},\boldsymbol{y}_0^{i-1}) & \, = \, & \mathcal{N}( \bunderline{A}^{-1}\bunderline{z}_i ; \bunderline{A}^{-1}\bunderline{\mu}_\epsilon + \bunderline{\mu}_\alpha , \bunderline{A}^{-1}\Sigma_{a}\bunderline{A}^{-T} - \nonumber \\
& & \, \, \, \, \, \,  \, \, \, \, \, \,  \bunderline{A}^{-1}\Sigma'_\epsilon\bunderline{A}^{-T} + \Sigma_\alpha ) , \nonumber \\
& \, = \, & \mathcal{N}\left( \bunderline{z}_i ; \bunderline{\mu}_\epsilon + \bunderline{A}\bunderline{\mu}_\alpha , \Sigma_{a} - \Sigma'_\epsilon + \bunderline{A} \Sigma_\alpha \bunderline{A}^T \right) , \nonumber \\
\label{2app75}
& \, = \, & \mathcal{N}\left( \bunderline{z}_i ; \bunderline{\mu}_i , \Sigma_i - \Sigma_\epsilon \right),
\end{IEEEeqnarray}
where $\bunderline{\mu}_i$ is defined later in (\ref{2app190}), and by performing substitutions from (\ref{fr50}) and (\ref{2app70}):
\begin{IEEEeqnarray}{rcl}
\label{2app150}
\Sigma_\epsilon & \, = \, & \Sigma'_\epsilon + \Sigma'''''_\epsilon, \\
\Sigma_i & \, = \, & \Sigma_{a} + |a|^2(\Sigma_{i-1}^{-1}+|x_{i-1}|^{2} \Sigma_{n}^{-1})^{-1}, \nonumber \\
\label{2app160}
& \, = \, & (1-|a|^2)\Sigma_z + |a|^2(\Sigma_{i-1}^{-1}+|x_{i-1}|^{2} \Sigma_{n}^{-1})^{-1}.
\end{IEEEeqnarray}

Noticing that if $\Sigma_{i-1}$ is proportional to the identity, then so is $\Sigma_i$, let:
\begin{equation}
\label{2app170}
\Sigma_i = \begin{bmatrix} \sigma_i^2 & 0 \\ 0 & \sigma_i^2 \end{bmatrix},
\end{equation}
where:
\begin{equation}
\label{2app180}
\sigma^2_i = (1-|a|^2)\sigma^2_z + |a|^2(\sigma_{i-1}^{-2}+|x_{i-1}|^{2} \sigma_{n}^{-2})^{-1}.
\end{equation}

Consider that the overall distribution of $z$ must be preserved, regardless of the input and noise, therefore:
\begin{equation}
\label{2app190}
\bunderline{\mu}_i \sim \mathcal{N}(\ubar{0}, \Sigma_z- (\Sigma_i-\Sigma_\epsilon)).
\end{equation}

To prove the final part of Lemma~5.4, i.e., that $\sigma_z^2 \geq \sigma_i^2$, consider again proof by induction. From (\ref{2app10}) it is known that $\sigma_0^2 = \sigma_z^2$, and thus consider (\ref{2app180}):
\begin{IEEEeqnarray}{rcl}
%\label{2app200}
\sigma_z^2 & \, \geq \, & \sigma^2_{i-1}, \nonumber \\
%\label{2app210}
& \, \geq \, & (\sigma_{i-1}^{-2}+|x_{i-1}|^{2} \sigma_{n}^{-2})^{-1}, \nonumber \\
& \, \geq \, & (1-|a|^2)\sigma^2_z + |a|^2(\sigma_{i-1}^{-2}+|x_{i-1}|^{2} \sigma_{n}^{-2})^{-1}, \nonumber \\
\label{2app220}
& \, = \, & \sigma^2_i.
\end{IEEEeqnarray}

\subsection*{Lemma 5}

For identity matrix, $I$, and PDS matrix $D$, there exists a PDS matrix $D'$ such that:
\begin{equation}
\label{3app20}
(I-D)^{-1}  =  I + D'.
\end{equation}

\subsection*{Proof:}

From [\cite{searle}~pp.~151]:
\begin{equation}
(I-D)^{-1} = I +(I-D)^{-1} D.
\end{equation}
Given that $(I-D)$ is a PDS matrix, $(I-D)^{-1}$ is also a PDS matrix. Also, since $D$ is a PDS matrix, then $(I-D)^{-1} D$ must be a PDS matrix, which is renamed $D'$ to prove Lemma~5.

\subsection*{Lemma 6}

For identity matrix, $I$, and PDS matrix $D$, there exists a PDS matrix $D'$ such that:
\begin{equation}
\label{3app10}
(I+D)^{-1}  =  I - D'. 
\end{equation}

\subsection*{Proof:}

From [\cite{searle}~pp.~151]:
\begin{equation}
\label{3app30}
(I+D)^{-1} = I - (I+D)^{-1} D.
\end{equation}
Given that $(I+D)$ is a PDS matrix, $(I+D)^{-1}$ is also a PDS matrix. Also, since $D$ is a PDS matrix, then $(I+D)^{-1} D$ must be a PDS matrix, which is renamed $D'$ to prove Lemma~6.

\subsection*{Lemma 7}

For $2 \times 2$ identity matrix, $I$, and $2 \times 2$ PDS matrix, $D$, with $(I-D)$ also a $2 \times 2$ PDS matrix, it follows that:
\begin{equation}
\label{4app10}
|I-D|  < 1.
\end{equation}

\subsection*{Proof:}
Let:
\begin{equation}
\label{4app20}
D = \begin{bmatrix} d_1 & d_2 \\ d_2 & d_4 \end{bmatrix},
\end{equation}
therefore:
\begin{IEEEeqnarray}{rcl}
|I-D| & \, = \, & (1-d_1)(1-d_4) - d_2^2, \nonumber \\
\label{4app30}
& \, = \, & 1 - d_1 -d_4 + d_1 d_4 -d_2^2,
\end{IEEEeqnarray}
consider $0<d_1,d_4<1$, therefore:
\begin{IEEEeqnarray}{rrcl}
\label{4app40}
 & d_1, d_4 & \, > \, & d_1 d_4, \\
\implies \, & \, 1 & \, >  \, & 1 - d_1 -d_4 + d_1 d_4 -d_2^2 \nonumber \\
\label{4app50}
& & \, = \, & |I-D|,
\end{IEEEeqnarray}
thus proving Lemma~7.

%\section{temp}
%\input{chmod2}


%\section{}

\section*{Acknowledgment}

The authors would like to thank Tobi Koch and Jossy Sayir for generously lending their expertise in a series of introductory discussions. Also to Edward Mottram for his help with the Lemma's, and to Ramji Venkataramanan for proof reading and internally reviewing the paper.

%, especially with Theorem~2.

% and Edward Mottram for his help with Lemma 4.\\

\ifCLASSOPTIONcaptionsoff
  \newpage
\fi



%\bibliographystyle{IEEEtran}
%\begin{thebibliography}{1}


%\end{thebibliography}

\bibliographystyle{ieee}
\bibliography{biblio}

\begin{thebibliography}{10}
\providecommand{\url}[1]{#1}
\csname url@samestyle\endcsname
\providecommand{\newblock}{\relax}
\providecommand{\bibinfo}[2]{#2}
\providecommand{\BIBentrySTDinterwordspacing}{\spaceskip=0pt\relax}
\providecommand{\BIBentryALTinterwordstretchfactor}{4}
\providecommand{\BIBentryALTinterwordspacing}{\spaceskip=\fontdimen2\font plus
\BIBentryALTinterwordstretchfactor\fontdimen3\font minus
  \fontdimen4\font\relax}
\providecommand{\BIBforeignlanguage}[2]{{%
\expandafter\ifx\csname l@#1\endcsname\relax
\typeout{** WARNING: IEEEtran.bst: No hyphenation pattern has been}%
\typeout{** loaded for the language `#1'. Using the pattern for}%
\typeout{** the default language instead.}%
\else
\language=\csname l@#1\endcsname
\fi
#2}}
\providecommand{\BIBdecl}{\relax}
\BIBdecl

\bibitem{bello}
P.~Bello, ``Characterization of randomly time-variant linear channels,''
  \emph{Communications Systems, IEEE Transactions on}, vol.~11, no.~4, pp.
  360--393, 1963.

\bibitem{durisi1}
G.~Durisi, U.~Schuster, H.~Bolcskei, and S.~Shamai, ``Noncoherent capacity of
  underspread fading channels,'' \emph{Information Theory, IEEE Transactions
  on}, vol.~56, no.~1, pp. 367--395, 2010.

\bibitem{us2}
\BIBentryALTinterwordspacing
S.~J. Herbert, I.~Wassell, T.-H. Loh, and J.~Rigelsford, ``Characterising the
  spectral properties and time variation of the in-vehicle wireless
  communication channell,'' \emph{Accepted for publication in IEEE Transactions
  on Communications}, 2014. [Online]. Available:
  \url{http://www.cl.cam.ac.uk/~sjh227/time\_var.pdf}
\BIBentrySTDinterwordspacing

\bibitem{wsnv}
J.~Dawson, D.~Hope, M.~Panitz, and C.~Christopoulos, ``Wireless networks in
  vehicles,'' in \emph{Electromagnetic Propagation in Structures and Buildings,
  2008 IET Seminar on}, dec. 2008, pp. 1 --6.

\bibitem{us1}
S.~Herbert, T.~Loh, and I.~Wassell, ``An impulse response model and $q$ factor
  estimation for vehicle cavities,'' \emph{IEEE Transactions on Vehicular
  Technology}, vol.~PP, no.~99, pp. 1--1, 2013.

\bibitem{durisi3}
\BIBentryALTinterwordspacing
G.~Durisi, V.~I. Morgenshtern, H.~Bolcskei, U.~G. Schuster, and
  S.~Shamai~(Shitz), \emph{Information theory of underspread {WSSUS} channels},
  2011, pp. 65--116. [Online]. Available:
  \url{http://www.nari.ee.ethz.ch/commth/pubs/p/dmbss\_book10}
\BIBentrySTDinterwordspacing

\bibitem{gallager}
R.~G. Gallager, \emph{Information Theory and Reliable Communication}.\hskip 1em
  plus 0.5em minus 0.4em\relax New York, NY, USA: John Wiley \& Sons, Inc.,
  1968.

\bibitem{ken69}
R.~S. Kennedy, \emph{Fading Dispersive Communication Channels}.\hskip 1em plus
  0.5em minus 0.4em\relax Wiley, 1969.

\bibitem{pie66}
J.~Pierce, ``Ultimate performance of m -ary transmissions on fading channels,''
  \emph{Information Theory, IEEE Transactions on}, vol.~12, no.~1, pp. 2--5,
  Jan 1966.

\bibitem{vit67}
A.~Viterbi, ``Performance of an m -ary orthogonal communication system using
  stationary stochastic signals,'' \emph{Information Theory, IEEE Transactions
  on}, vol.~13, no.~3, pp. 414--422, July 1967.

\bibitem{shannon1948}
\BIBentryALTinterwordspacing
C.~E. Shannon, ``A mathematical theory of communication,'' \emph{The Bell
  System Technical Journal}, vol.~27, pp. 379--423, 623--656, July, October
  1948. [Online]. Available:
  \url{http://cm.bell-labs.com/cm/ms/what/shannonday/shannon1948.pdf}
\BIBentrySTDinterwordspacing

\bibitem{durisi2}
G.~Durisi, H.~Bolcskei, and S.~Shamai, ``Capacity of underspread wssus fading
  channels in the wideband regime,'' in \emph{Information Theory, 2006 IEEE
  International Symposium on}, 2006, pp. 1500--1504.

\bibitem{tse1}
D.~Tse and P.~Viswanath, \emph{Fundamentals of wireless communication}.\hskip
  1em plus 0.5em minus 0.4em\relax New York, NY, USA: Cambridge University
  Press, 2005.

\bibitem{goldsmith}
A.~Goldsmith, \emph{Wireless Communications}.\hskip 1em plus 0.5em minus
  0.4em\relax Cambridge University Press, 2005.

\bibitem{kalman}
\BIBentryALTinterwordspacing
R.~E. Kalman, ``A new approach to linear filtering and prediction problems,''
  1960. [Online]. Available:
  \url{http://www.cs.unc.edu/~welch/kalman/media/pdf/Kalman1960.pdf}
\BIBentrySTDinterwordspacing

\bibitem{sair}
J.~Andersen, K.~L. Chee, M.~Jacob, G.~Pedersen, and T.~Kurner, ``Reverberation
  and absorption in an aircraft cabin with the impact of passengers,''
  \emph{Antennas and Propagation, IEEE Transactions on}, vol.~60, no.~5, pp.
  2472 --2480, may 2012.

\bibitem{kl}
T.~Koch and A.~Lapidoth, ``On multipath fading channels at high snr,''
  \emph{Information Theory, IEEE Transactions on}, vol.~56, no.~12, pp.
  5945--5957, 2010.

\bibitem{zigbee}
\BIBentryALTinterwordspacing
 [Online]. Available: \url{http://www.zigbee.org/}
\BIBentrySTDinterwordspacing

\bibitem{searle}
S.~R. Searle, \emph{Matrix Algebra Useful for Statistics}.\hskip 1em plus 0.5em
  minus 0.4em\relax New York, NY, USA: John Wiley \& Sons, Inc., 1982.

\end{thebibliography}

%\begin{biography}[{\includegraphics[width=1in,height=1.25in,clip,keepaspectratio]{mshell}}]{Michael Shell}

%\end{IEEEbiography}


%\newpage


\begin{IEEEbiographynophoto}{Steven Herbert}%[{\includegraphics[width=1in,height=1.25in,clip,keepaspectratio]{sjh227.jpg}}]
(S'12) received the B.A. degree [subsequently promoted to M.A. (Cantab) in 2013] from the University of Cambridge, Cambridge, U.K., in 2010 and the M.Eng. degree. He is currently working towards the Ph.D. degree with the University of Cambridge.\\
He is currently with the Digital Technologies Group, Computer Laboratory, University of Cambridge, and the Electromagnetic Technologies Group, National Physical Laboratory, Middlesex, U.K.
\end{IEEEbiographynophoto}

\begin{IEEEbiographynophoto}{Ian Wassell}%[{\includegraphics[width=1in,height=1.25in,clip,keepaspectratio]{ijw_head_shot.jpg}}]
received his B.Sc. and B.Eng. degrees from the University of Loughborough in 1983, and his Ph.D. degree from the University of Southampton in 1990. He is a Senior Lecturer at the University of Cambridge Computer Laboratory and has in excess of 15 years experience in the simulation and design of radio communication systems gained via a number of positions in industry and higher education. He has published more than 190 papers concerning wireless communication systems and his current research interests include: fixed wireless access, sensor networks, cooperative networks, propagation modelling, compressive sensing and cognitive radio. He is a member of the IET and a Chartered Engineer.
\end{IEEEbiographynophoto}


\begin{IEEEbiographynophoto}{Tian Hong Loh} %[{\includegraphics[width=1in,height=1.25in,clip,keepaspectratio]{tian.jpg}}]
(S'03-M'05) was born in Johor, Malaysia. He received the B.Eng. degree (first class) from Nottingham Trent University, Nottingham, U.K., and the Ph.D. degree from the University of Warwick, Coventry, U.K., in 1999 and 2005, respectively, both in electrical and electronic engineering. He joined the National Physical Laboratory, Teddington, U.K., in 2005 as a Higher Research Scientist and since 2009 he has been a Senior Research Scientist, involved in work on fundamental research and develop measurement technologies in support of the electronics and communication industry. Since 2011, he has been appointed as RF and Microwave technical theme leader, involved in physical programme formulation and strategy development. His current research interests include metamaterials, computational electromagnetics, small antenna, smart antennas, multiple-input-multiple-output antennas, wireless sensor networks.
\end{IEEEbiographynophoto}

%\begin{IEEEbiography}[{\includegraphics[width=1in,height=1.25in,clip,keepaspectratio]{JMR_Profile_pic2.png}}]{Jonathan Rigelsford}
%(M'05, SM'13) received the M.Eng. and Ph.D. degrees in electronic engineering from the University of Hull, Hull, U.K. in 1997 and 2001, respectively. From 2000 to 2002, he worked as a Senior Design Engineer at Jaybeam Limited, designing antennas for cellular base stations. Since late 2002, he has been a Senior Experimental Officer for the Communication Group within the Department of Electronic and Electrical Engineering, University of Sheffield, Sheffield, U.K. He has been an active member of the Antenna Interface Standards Group (AISG) from 2002 to 2010 being elected to the board of directors during that time. More recently, he has become the Secretary to the Wireless Friendly Building Forum, an industrial/academic initiative to promote understanding of radio propagation within the built environment.
%\end{IEEEbiography}

%\newpage


%\vfill


\end{document}